\documentclass[11pt,a4paper]{article}
\usepackage{jheppub}
\usepackage{amsmath,amssymb,amsfonts}
\usepackage{graphicx,graphics}
\usepackage{graphics}
\usepackage{url}
\usepackage{slashed}
\usepackage{color}
\usepackage{mathtools}
\usepackage[utf8]{inputenc}
\usepackage{subcaption}

\usepackage{siunitx}
\newcommand{\msbar}{\overline{\mathrm{MS}}}
\newcommand{\renorm}{\mu_{\text{ren}}}
\newcommand{\fact}{\mu_{\text{fact}}}
\newcommand{\frag}{\mu_{\text{frag}}}
\newcommand{\dd}{\mathrm{d}}

\DeclareMathOperator{\sub}{sub}

\title{Improving the description of dimuon production in neutrino-nucleus collisions using the SACOT-$\chi$ scheme}

\affiliation{University of Jyväskylä, Department of Physics, P.O. Box 35, FI-40014 University of Jyväskylä, Finland}
\affiliation{Helsinki Institute of Physics, P.O. Box 64, FI-00014 University of Helsinki, Finland}

\emailAdd{ilkka.m.helenius@jyu.fi}
\emailAdd{hannu.paukkunen@jyu.fi}
\emailAdd{sami.a.yrjanheikki@jyu.fi}

\abstract{
Dimuon production in deeply inelastic scattering between neutrinos and nuclei plays an important role in constraining the strange-quark parton distribution functions (PDFs). Here, we present a self-contained calculation of this process consisting of a next-to-leading order semi-inclusive charmed-hadron production in the SACOT-$\chi$ general-mass variable-flavor-number scheme, followed by a semi-leptonic decay of the charmed hadron. We find that invoking the SACOT-$\chi$ scheme introduces modifications up to $\SI{20}{\percent}$ in comparison to our previous esimates, where only kinematic mass effects were considered through the slow-rescaling variable. We reiterate our earlier observation that the effective acceptance correction --- typically used in global PDF fits as a simplifying approximation --- depends on the perturbative order, PDFs, scales, and also on the treatment of heavy-quark effects. We find good agreement with the corresponding data from the NuTeV experiment.
}


\begin{document}

\author{Ilkka Helenius,}
\author{Hannu Paukkunen and}
\author{Sami Yrjänheikki}

\maketitle

\section{Introduction}

Parton distribution functions (PDFs) \cite{Kovarik:2019xvh,Ethier:2020way,Klasen:2023uqj} are a necessary non-perturbative input for perturbative Quantum Chromodynamics (QCD) calculations involving colliding hadrons, making them especially important for physics at the Large Hadron Collider (LHC) and the future Electron-Ion Collider (EIC) \cite{AbdulKhalek:2021gbh}. For processes involving hadronization, fragmentation functions (FFs) \cite{Metz:2016swz} provide a similar non-perturbative input on the final-state side. Both are universal and process-independent in the collinear factorization framework \cite{Collins:1989gx}. While neither is fully perturbatively calculable, their scale evolution is predicted by the Dokshitzer-Gribov-Lipatov-Altarelli-Parisi (DGLAP) evolution equations \cite{Gribov:1972ri,Gribov:1972rt,Dokshitzer:1977sg,Altarelli:1977zs}. 

Owing to the non-perturbative nature of PDFs, the initial condition that can be evolved with DGLAP is determined in a global analysis, which aims to include as much data as possible to simultaneously constrain the PDFs with multiple complimentary processes. The strange-quark distribution is one of the most poorly constrained flavor components. Major sources of constraining power include semi-inclusive dimuon production in neutrino-nucleus collisions \cite{CCFR:1994ikl,NuTeV:2001dfo,NuTeV:2007uwm,CHORUS:2008vjb,NOMAD:2013hbk}, $W$ and $Z$ boson production in proton-proton collisions \cite{Kusina:2012vh}, and semi-inclusive kaon production in charged-lepton-proton collisions \cite{HERMES:2013ztj,Borsa:2017vwy,Sato:2019yez}. The large interaction scale $Q^2\sim \SI{e4}{GeV^2}$ in $W$ and $Z$ boson production obscures the underlying non-perturbative strange-quark distribution, as most strange quarks at such high scales are perturbatively generated via gluon splittings $g\to s\overline s$. As for kaon production, its precision is limited by the uncertainties of the kaon FFs. Due to these challenges, dimuon production still remains an important constraint for strangeness.

Dimuon production proceeds through a charged-current deep inelastic scattering (DIS) that produces a charm quark in the final state. Due to the hierarchy in the Cabibbo-Kobayashi-Maskawa (CKM) matrix elements \cite{ParticleDataGroup:2024cfk}, this underlying partonic process predominantly probes the strange-quark parton distribution. The produced charm will then hadronize into a charmed hadron (mostly a $D$ meson), which eventually decays semimuonically. To consistently use such measurements in a global analysis, the hadronization and decay must be included as parts of the calculation. Even still, the typical approach used in many PDF fits \cite{Alekhin:2017kpj,Hou:2019efy,Bailey:2020ooq,NNPDF:2021njg} is to assume a factorization-like connection between charm and dimuon production:
\begin{equation}
\label{eq:charm_prod_factorization}
    \dd\sigma(\nu N\to \mu\mu X)=\mathcal{A}\mathcal{B}_\mu \dd\sigma(\nu N\to c\mu X),
\end{equation}
where $\dd\sigma(\nu N\to \mu\mu X)$ is the dimuon production cross section and $\dd\sigma(\nu N\to c\mu X)$ is the inclusive charm production cross section. The acceptance $\mathcal{A}$ takes the experimentally-imposed energy cut on the decay muon into account, and $\mathcal{B}_\mu\sim 0.09$ is the semileptonic branching ratio.

Equation~\eqref{eq:charm_prod_factorization} holds only in the leading-order (LO) fixed-flavor number scheme (FFNS). Going to the next-to-leading order (NLO) or even to a LO variable-flavor number scheme (VFNS) calculation breaks this factorization. Regardless of the perturbative order or flavor number scheme, the acceptance $\mathcal A$ has to be taken as an input and is typically calculated separately with Monte-Carlo methods \cite{Mason:2006qa}. Beyond the LO FFNS calculation, the acceptance can still in principle be computed, but it then becomes PDF-, scale-, and scheme-dependent. In practice, almost every computation would need a dedicated acceptance correction, which is not practical for fitting. Additionally, it is not entirely clear what value should be assigned to $\mathcal{B}_\mu$ --- see section~4.6 in ref.~\cite{Helenius:2024fow} --- and it is usually given a $\SI{10}{\percent}$ normalization uncertainty even separately for neutrino and antineutrino scattering. A situation where an acceptance correction depends on e.g. PDFs, is also typical for LHC measurements. Such a correction is usually computed with several PDF sets and the resulting spread included in the systematic uncertainties. In the case of dimuon production, the modeling itself can be improved to eliminate the acceptance correction $\mathcal{A}$ entirely.

In our previous work \cite{Helenius:2024fow}, we introduced the semi-inclusive DIS (SIDIS) approach for computing dimuon production in neutrino-nucleus collisions. This approach, which we recapitulate in section~\ref{sec:sidis_approach}, foregoes eq.~\eqref{eq:charm_prod_factorization} in favor of computing charmed-hadron production and the subsequent muonic decay process directly. Previously, we implemented the charmed-hadron production in the zero-mass (ZM) and slow-rescaling (SR) schemes. In the ZM scheme, the charm quark is taken to be massless. In the SR scheme, also known as the intermediate-mass (IM) scheme \cite{Nadolsky:2009ge}, the charm-quark mass is taken into account only kinematically, which amounts to the prescription of replacing the Björken-$x$ in the convolution integrals by the slow-rescaling variable $\chi=x(1+m_c^2/Q^2)$. At leading order this prescription happens to be exact in the case of charged-current DIS, but there are additional mass corrections at higher orders.

In this paper, we expand on our previous work by implementing the charmed-hadron production in the simplified Aivazis-Collins-Olness-Tung (SACOT) general-mass variable-flavor number scheme (GM-VFNS) \cite{Olness:1987ep,Aivazis:1993kh,Aivazis:1993pi,Kretzer:1997pd,Collins:1998rz,Kramer:2000hn}. Specifically, we use the SACOT-$\chi$ variant \cite{Tung:2001mv,Kretzer:2003it,Guzzi:2011ew,Gao:2021fle,Risse:2025smp} where the slow-rescaling variable $\chi$ is used just as in the aforementioned SR scheme. The difference between the SACOT and SACOT-$\chi$ schemes is mostly relevant near the charm threshold region, since the SACOT-$\chi$ scheme respects the kinematic constraints of producing a heavy quark.

The purpose of heavy-quark mass schemes is to resum the logarithms $\log^k(Q^2/m^2)$ appearing in the massive coefficient functions. These logarithms, while small at $Q^2\lesssim m^2$, can become large when $Q^2\gg m^2$ and therefore spoil the convergence of the perturbative series. There is no unique approach to this and as a result, many different schemes exist in the literature. For example, the \texttt{CT18} \cite{Hou:2019efy} analysis uses the aforementioned SACOT-$\chi$ scheme, while the \texttt{MSHT20} PDF analysis \cite{Bailey:2020ooq} uses the Thorne-Roberts (TR) scheme \cite{Thorne:1997ga,Thorne:2006qt,Thorne:2012az}, and the \texttt{NNPDF4.0} \cite{NNPDF:2021njg} and \texttt{nNNPDF3.0} \cite{AbdulKhalek:2022fyi} analyses use the ``fixed-order plus next-to-leading logs'' (FONLL) scheme \cite{Cacciari:1998it,Forte:2010ta}. If one could sum the entire perturbative series, the differences between these schemes would vanish. Since this cannot presently be done, there will be differences among the heavy-quark schemes due to the truncation of differently-ordered perturbative series.

Currently, data on differential dimuon cross sections are available from the NuTeV \cite{NuTeV:2007uwm} and CCFR \cite{CCFR:1994ikl} experiments. These data are from fixed-target neutrino beam experiments on an iron target. In this study, we consider only the NuTeV data with neutrino and antineutrino beams, as the CCFR kinematics are comparable to NuTeV. However, future dimuon data are expected from the new far-forward experiments at LHC \cite{Cruz-Martinez:2023sdv,Ariga:2025qup}. The FASER \cite{FASER:2019dxq} and SND@LHC \cite{SNDLHC:2022ihg} experiments have already detected their first collider neutrinos and reported their first results \cite{FASER:2023zcr,FASER:2024hoe,FASER:2024ref,SNDLHC:2023mib,SNDLHC:2023pun}. These neutrinos are produced along the beamline in e.g. proton-proton collisions from the decays of light hadrons, and can reach energies of up to several TeV. FASER2 and AdvSND, as part of the proposed Forward Physics Facility (FPF) \cite{Feng:2022inv}, could provide further dimuon data to help constrain PDFs. In addition, the recently-approved SHiP experiment \cite{Alekhin:2015byh,SHiP:2025ows} at the SPS Beam Dump Facility is expected to deliver a new high-statistics dimuon data set.

\section{Dimuon production in the SIDIS approach}
\label{sec:sidis_approach}

\begin{figure}
    \centering
    \includegraphics{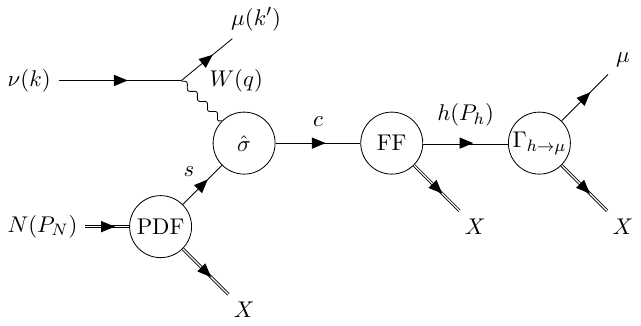}
    \caption{Diagrammatic depiction of dimuon production in neutrino-nucleus scattering. Relevant momenta are indicated in parentheses.}
    \label{fig:dimuon_prod}
\end{figure}

Figure~\ref{fig:dimuon_prod} shows the full dimuon production process, where an intermediate-state charmed hadron $h$ is produced and which then decays into a muon plus a hadronic state. As explained in detail in ref.~\cite{Helenius:2024fow} (see also e.g. ref.~\cite{Czakon:2022pyz}, which shares a similar idea), we can make the split between the hadron production and decay more explicit by using the narrow-width approximation to write the dimuon cross section as
\begin{equation}
    \frac{\dd\sigma(\nu N\to \mu\mu X)}{\dd x \, \dd y}=\sum_h\int \dd z \, \frac{\dd\sigma(\nu N\to \mu hX)}{\dd x \, \dd y \, \dd z} B_{h\to \mu}(E_h=zyE_\nu, E_\mu^{\min}),
\end{equation}
where $\dd\sigma(\nu N\to \mu hX)$ is the semi-inclusive cross section for the production of a hadron $h$ and $B_{h\to \mu}(z, E_\mu^{\min})$ is the energy-dependent branching ratio for the decay of the hadron $h$ to a muon. 

The semi-inclusive cross section is given by
\begin{equation}
\label{eq:sidis_hadron_cross_section}
\begin{aligned}
	\frac{\dd\sigma(\nu_\mu N\to \mu h X)}{\dd x \, \dd y \, \dd z}&= \frac{G_F^2M_W^4}{\left(Q^2+M_W^2\right)^2}
  \frac{Q^2}{2\pi xy}   
 \\ &\phantom{=} \ \times \bigg[xy^2 F_1(x, z, Q^2)+\left(1-y-\frac{xy M^2}{s-M^2}\right)F_2(x, z, Q^2) \\ &\phantom{= \times\bigg[} \ \pm xy\left(1-\frac{y}{2}\right)F_3(x, z, Q^2)\bigg],
\end{aligned}
\end{equation}
where $G_F\approx \SI{1.166379e-5}{GeV^{-2}}$ is the Fermi coupling constant and $M_W\approx \SI{80.377}{GeV}$ is the mass of the $W$ boson. The $(-)+$ sign in the $F_3$ term corresponds to (anti)neutrino scattering. The kinematic variables are the usual SIDIS variables:
\begin{equation}
\label{eq:sidis_kinematics}
\begin{aligned}
	Q^2&=-q^2 =- (k-k')^2 \geq 0 \,, \\
	x&=\frac{Q^2}{2P_N\cdot q} \,, \\
	y&=\frac{P_N\cdot q}{P_N\cdot k} \,, \\
    z&=\frac{P_N\cdot P_h}{P_N\cdot q} \,, \\
    s-M^2&=(k+P_N)^2-M^2=\frac{Q^2}{xy} \,,
\end{aligned}
\end{equation}
where $M$ is the proton mass and $k$, $k'$, $q$, $P_N$, and $P_h$ are the momenta of the incoming neutrino, outgoing muon, $W$ boson, target, and the hadron $h$.

The energy-dependent branching ratio takes the energy cut of the decay muon into account; in the limit $E_\mu^{\min}\to 0$, it reduces to the usual branching ratio. The energy-dependent branching ratio can be expressed in the rest frame of the nuclear target in terms of the decay function $d_{h\to \mu}$, a Lorentz-invariant scalar function, as
\begin{equation}
    B_{h\to\mu}(E_h, E_\mu^{\text{min}})=\frac{\pi}{m_h \Gamma_{\text{tot}}}\int \dd\rho \, \rho E_\mu^2\int \dd(\cos\theta) \, d_{h\to \mu}(w)\big|_{E_\mu=\rho E_h\geq E_\mu^{\text{min}}},
\end{equation}
where $m_h$ is the mass of $h$, $\Gamma_{\text{tot}}$ is the total decay width of $h$, $\rho=E_\mu/E_h$, $\theta$ is the relative angle between the momenta of the decay muon and $h$, and $w=(P_\mu\cdot P_h)/m_h^2$. The decay function is parametrized with the functional form
\begin{equation}
    d_{h\to\mu}(w)=Nw^\alpha(1-\gamma w)^\beta\Theta(0\leq w\leq 1/\gamma),
\end{equation}
where $N$, $\alpha$, $\beta$, and $\gamma$ are free parameters, and $\Theta$ is the Heaviside theta function. The decay function is then fitted to $D$-meson decay data obtained by CLEO \cite{CLEO:2006ivk} in $e^+ e^-$ collisions --- see details in ref. \cite{Helenius:2024fow}.

\section{Deep inelastic scattering in the SACOT-$\chi$ scheme}

In the SACOT-$\chi$ scheme, the freedom in the choice of the mass dependence in the cross section is used to simplify the calculation by taking an initial-state heavy quark or the parton fragmenting into a heavy-quark hadron as massless. Thus, we only need to consider the charm quark to be massive.\footnote{The contribution of the bottom quark is negligble due to the CKM matrix element $\left|V_{cb}\right|^2\sim 0.0017$.} The $\chi$ in the SACOT-$\chi$ scheme refers to the use of the slow-rescaling variable $\chi$ in the structure functions, instead of the plain momentum fraction $x$. The slow-rescaling variable depends on the number of final-state charm quarks $n$, and is defined as
\begin{equation}
    \chi_n=x\left(1+\frac{(n m_c)^2}{Q^2}\right).
\end{equation}
In most cases, the slow-rescaling variable is $\chi_1$ and for convenience, we denote $\chi\equiv\chi_1$.

\subsection{Semi-inclusive DIS}

The fixed-order semi-inclusive structure functions $F_i$ ($i=1,2,3$) are \cite{Gluck:1997sj}
\begin{equation}
\label{eq:sidis_unsub_sf}
\begin{aligned}
    F_i(x, z, Q^2, m_c^2)&=\sum_q \kappa_i \left|V_{qc}\right|^2 q(\fact^2)\otimes \left[H_{\text{LO}}^{qq}+\frac{\alpha_s(\renorm^2)}{2\pi}H_i^{qq}(\fact^2, m_c^2)\right]\otimes D_{c\to h} \\
        &\phantom{=} \ + \kappa_i g(\fact^2)\otimes \left[\frac{\alpha_s(\renorm^2)}{2\pi}H_i^{qg}(\fact^2, m_c^2)\right]\otimes D_{c\to h},
\end{aligned}
\end{equation}
where $V_{qc}$ are the CKM matrix elements. The renormalization and factorization scales are denoted by $\renorm^2$ and $\fact^2$, respectively. The double-convolution is defined as
\begin{equation}
\label{eq:double_conv_def}
    q\otimes H \otimes D\equiv \int_\chi^1 \frac{\dd\xi}{\xi}\int_{\max(z,\zeta_{\min})}^1\frac{\dd\zeta}{\zeta}q(\chi/\xi)H(\xi, \zeta)D(z/\zeta),
\end{equation}
with
\begin{equation}
    \zeta_{\min}=\frac{(1-\lambda)\xi}{1-\lambda\xi} \quad\text{where}\quad \lambda=\frac{Q^2}{Q^2+m_c^2}.
\end{equation}
The limits of integration in the double-convolution arise from the hard partonic cross section, which also contains all the dependence on the charm mass through the variables $\chi$ and $\lambda$. For practical purposes, however, we define the convolution in eq.~\eqref{eq:double_conv_def} as in ref.~\cite{Gluck:1997sj}. The definition of the relation between the final-state hadronic and partonic momenta is not unique \cite{Kniehl:2015fla}, but all such definitions must coincide with each other in the limit $m_c^2/Q^2\to 0$. Ref.~\cite{Gluck:1997sj}, and therefore eq.~\eqref{eq:sidis_unsub_sf}, uses the ratio of partonic and hadronic energies in the target-rest frame.

At leading order,
\begin{equation}
    H_{\text{LO}}^{qq}(\xi, \zeta)=\delta(1-\xi)\delta(1-\zeta).
\end{equation}
This is the same as in the massless case, as there are no leading-order mass effects at the matrix-element level. The hard NLO coefficients $H_i^{qq}$ and $H_i^{gq}$ can be found in ref.~\cite{Gluck:1997sj}. The normalizations of the structure functions are given by
\begin{equation}
\label{eq:sidis_normalizations}
    \kappa_1=1, \quad \kappa_2=2\chi, \quad \kappa_3=\pm 2.
\end{equation}
The sign of $\kappa_3$ is negative when antiquarks are involved (i.e. in the case of an antineutrino beam). For the fragmentation functions, we use the GM-VFNS NLO sets \texttt{kkks08} (specifically, the OPAL set) \cite{Kneesch:2007ey} for $D^0$ and $D^+$ and \texttt{bkk05} \cite{Kniehl:2006mw} for $D_s$ and $\Lambda_c^+$.

Equation~\eqref{eq:sidis_unsub_sf} is valid for $Q^2\lesssim m_c^2$. Notably, the fragmentation function is scale-independent in that case. When $Q^2\gg m_c^2$, the mass logarithms $\log(Q^2/m_c^2)$ are large and need to be resummed into the scale evolution of the fragmentation function. This resummation to all orders is implemented by the DGLAP evolution equations. However, in order to take care of the double-counting with the $\mathcal{O}(\alpha_s)$ gluon-radiation processes, it is sufficient to consider only the first $\mathcal{O}(\alpha_s)$ correction of the full DGLAP-evolved fragmentation function. This truncated expansion of the fragmentation function $D_{c\to h}$ depending on the fragmentation scale $\frag^2$ in the $\msbar$ scheme can be written as \cite{Kretzer:1998nt}
\begin{equation}
\label{eq:scale_dep_ff}
\begin{aligned}
    D_{c\to h}(z, \frag^2)&=D_{c\to h}(z)+\frac{\alpha_s(\renorm^2)}{2\pi}C_F\int_z^1 \frac{\dd \zeta}{\zeta} D_{c\to h}(z/\zeta) \\
    &\phantom{=} \ \times\left[\frac{1+\zeta^2}{1-\zeta}\left(\log\frac{\frag^2}{m_c^2}-1-2\log(1-\zeta)\right)\right]_++\mathcal{O}(\alpha_s^2).
\end{aligned}
\end{equation}
The scale-independent fragmentation function in eq.~\eqref{eq:sidis_unsub_sf} can be replaced with the scale-dependent one given in eq.~\eqref{eq:scale_dep_ff} by inverting the relationship in eq.~\eqref{eq:scale_dep_ff} as
\begin{equation}
\begin{aligned}
    D_{c\to h}(z)&=D_{c\to h}(z, \frag^2)-\frac{\alpha_s(\renorm^2)}{2\pi}C_F\int_z^1 \frac{\dd \zeta}{\zeta} D_{c\to h}(z/\zeta, \frag^2) \\
    &\phantom{=} \ \times\left[\frac{1+\zeta^2}{1-\zeta}\left(\log\frac{\frag^2}{m_c^2}-1-2\log(1-\zeta)\right)\right]_++\mathcal{O}(\alpha_s^2)
\end{aligned}
\end{equation}
and inserting this back into eq.~\eqref{eq:sidis_unsub_sf}. This results in
\begin{equation}
\label{eq:sidis_sub_sf}
\begin{aligned}
    F_i(x, z, Q^2, m_c^2)&=\sum_{q}\kappa_i \left|V_{qc}\right|^2 q(\fact^2)\otimes \left[H_{\text{LO}}^{qq}+\frac{\alpha_s(\renorm^2)}{2\pi}H_i^{qq}(\fact^2, m_c^2)\right]\otimes D_{c\to h}(\frag^2) \\
    &\phantom{=} \ + \kappa_i g(\fact^2)\otimes \left[\frac{\alpha_s(\renorm^2)}{2\pi}H_i^{qg}(\fact^2, m_c^2)\right]\otimes D_{c\to h}(\frag^2) \\
    &\phantom{=} \ - \sum_q\kappa_i \left|V_{qc}\right|^2 \sub_i^{qc}(x, z, \renorm^2, \fact^2, \frag^2, m_c^2),
\end{aligned}
\end{equation}
where the subtraction term is given by
\begin{equation}
\label{eq:sidis_D_sub_term}
\begin{aligned}
    \sub_i^{qc}(x, z, \renorm^2, \fact^2, \frag^2, m_c^2)&=\frac{\alpha_s(\renorm^2)}{2\pi}q(\chi, \fact^2)\int_z^1 \frac{\dd \zeta}{\zeta} D_{c\to h}(z/\zeta, \frag^2) \\
    &\phantom{=} \ \times C_F\left[\frac{1+\zeta^2}{1-\zeta}\left(\log\frac{\frag^2}{m_c^2}-1-2\log(1-\zeta)\right)\right]_+.
\end{aligned}
\end{equation}
Besides the term $(\text{splitting function})\times (\text{mass logarithm})$, eq.~\eqref{eq:sidis_D_sub_term} includes a finite term. This scheme-dependent term ensures that the $\msbar$ coefficient function is recovered in the massless limit, i.e.
\begin{equation}
    \lim_{m_c\to 0}F_i(x, z, Q^2, m_c^2)=F_i^{\msbar}(x, z, Q^2).
\end{equation}

Equation~\eqref{eq:sidis_sub_sf} includes the quark-to-quark and gluon-to-quark channels where the fragmenting parton is a charm quark, but not gluon fragmentation or quark fragmentation where the fragmenting quark is not a charm quark. Both channels involve a gluon splitting $g\to c\overline c$, either directly (gluon fragmentation) or indirectly via an emitted gluon from a quark line $q'\to q'g\to q'c\overline c$ (quark fragmentation of a non-charm quark), and are thus beyond-$\alpha_s$ corrections. They can therefore also be included in the structure functions, as a structure function with and without them are formally equivalent at order $\alpha_s$. In both cases, the respective coefficient function is computed in the SACOT-$\chi$ scheme with the massless coefficient, which can be found e.g. in ref.~\cite{Furmanski:1981cw}. In addition, the correct rescaling variable is $\chi_2=x(1+4m_c^2/Q^2)$, again due to the splitting $g\to c\overline c$. These channels are included only for completeness as their contributions are negligble --- see figure 6 in ref.~\cite{Helenius:2024fow} --- and are not explicitly given in eq.~\eqref{eq:sidis_sub_sf}.

\subsection{Inclusive DIS}
\label{sec:inclusive_dis_sacot}

While the inclusive DIS calculation is not necessary for dimuon production in our framework, it is still useful for comparing to other works. To do this, we also need to implement inclusive charm production in DIS in the SACOT-$\chi$ scheme. 

The fixed-order structure functions are given by
\begin{equation}
\begin{aligned}
    F_i(x, Q^2, m_c^2)&=\sum_{q} \kappa_i \left|V_{qc}\right|^2 q(\fact^2)\otimes \left[H_{\text{LO}}^q +\frac{\alpha_s(\renorm^2)}{2\pi}H_i^q(\fact^2, m_c^2)\right] \\
    &\phantom{=} \ + \kappa_i g(\fact^2)\otimes \left[\frac{\alpha_s(\renorm^2)}{2\pi}H_i^g(\fact^2, m_c^2)\right]
\end{aligned}
\end{equation}
where the convolution is defined as
\begin{equation}
    q\otimes H\equiv \int_\chi^1 \frac{\dd\xi}{\xi}q(\chi/\xi) H(\xi).
\end{equation}
The LO coefficient is given by
\begin{equation}
    H_{\text{LO}}^q(\xi)=\delta(1-\xi),
\end{equation}
while the NLO coefficients $H_i^q$ and $H_i^g$ can be found in ref.~\cite{Gluck:1996ve}. In this case, a subtraction term is needed to regulate the logarithms $\log\left(Q^2/m_c^2\right)$ associated with the gluon-initiated channels above the charm-mass threshold $Q^2>m_c^2$. The subtraction term is \cite{Aivazis:1993pi,Kretzer:1998ju}
\begin{equation}
    \sub_i^g(x, \renorm^2, \fact^2, m_c^2)=\frac{\alpha_s(\renorm^2)}{2\pi}\log\left(\frac{\fact^2}{m_c^2}\right) g(\fact^2) \otimes P_{qg},
\end{equation}
where $P_{qg}(\xi)=\frac{1}{2}(\xi^2+(1-\xi)^2)$. This time, the subtraction term is absorbed into the anticharm-quark distribution
\begin{equation}
    \overline c(\chi, \fact^2)=\frac{\alpha_s(\renorm^2)}{2\pi}\log\left(\frac{\fact^2}{m_c^2}\right)g(\fact^2)\otimes P_{qg}+\mathcal{O}(\alpha_s^2),
\end{equation}
which induces a charm-initiated channel.\footnote{For antineutrino scattering, the subtraction term is absorbed into the charm-quark distribution.}

In our implementation, we also include the massless NLO coefficient $C_i^q$ in the charm-initiated channels. Since there are no massive final-state quarks in these charm-initiated channels, the structure function $F_2$ will be proportional to just $x$ instead of $\chi$ in the SACOT scheme. Nevertheless, in keeping with the SACOT-$\chi$ scheme, the convolutions between PDFs and coefficient functions will still use the slow-rescaling variable $\chi$. Thus, the subtracted structure functions read
\begin{equation}
\label{eq:inclusive_dis_sacot}
\begin{aligned}
    F_i(x, Q^2, m_c^2)&=\sum_q \kappa_i \left|V_{qc}\right|^2 q(\fact^2)\otimes \left[H_{\text{LO}}^q +\frac{\alpha_s(\renorm^2)}{2\pi} H_i^q(\fact^2, m_c^2)\right] \\
    &\phantom{=} \ +\kappa_i\frac{\alpha_s(\renorm^2)}{2\pi}g(\fact^2)\otimes H_i^g(\fact^2, m_c^2) \\
    &\phantom{=} \ - \tilde{\kappa}_i\frac{\alpha_s(\renorm^2)}{2\pi}\log\left(\frac{\fact^2}{m_c^2}\right)g(\fact^2)\otimes P_{qg} \\
    &\phantom{=} \ + \sum_{\overline q'} \tilde{\kappa}_i \left|V_{cq'}\right|^2 \overline c(\fact^2)\otimes \left[H_{\text{LO}}^q+\frac{\alpha_s(\renorm^2)}{2\pi}C_i^q(\fact^2)\right],
\end{aligned}
\end{equation}
where
\begin{equation}
\begin{aligned}
    &\kappa_1=1, \quad \kappa_2=2\chi, \quad \kappa_3=\pm 2, \\ 
    &\tilde{\kappa}_1=1, \quad \tilde{\kappa}_2=2x, \quad \tilde{\kappa}_3=\pm 2.
\end{aligned}
\end{equation}

\section{Results}

\begin{figure}
    \centering
    \includegraphics[width=\linewidth]{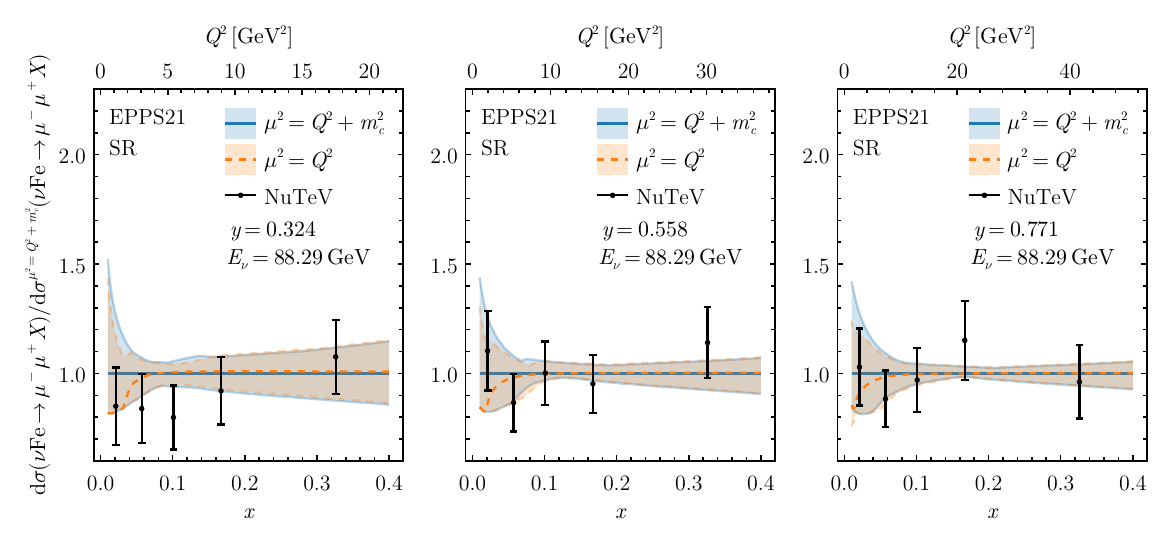}
    \includegraphics[width=\linewidth]{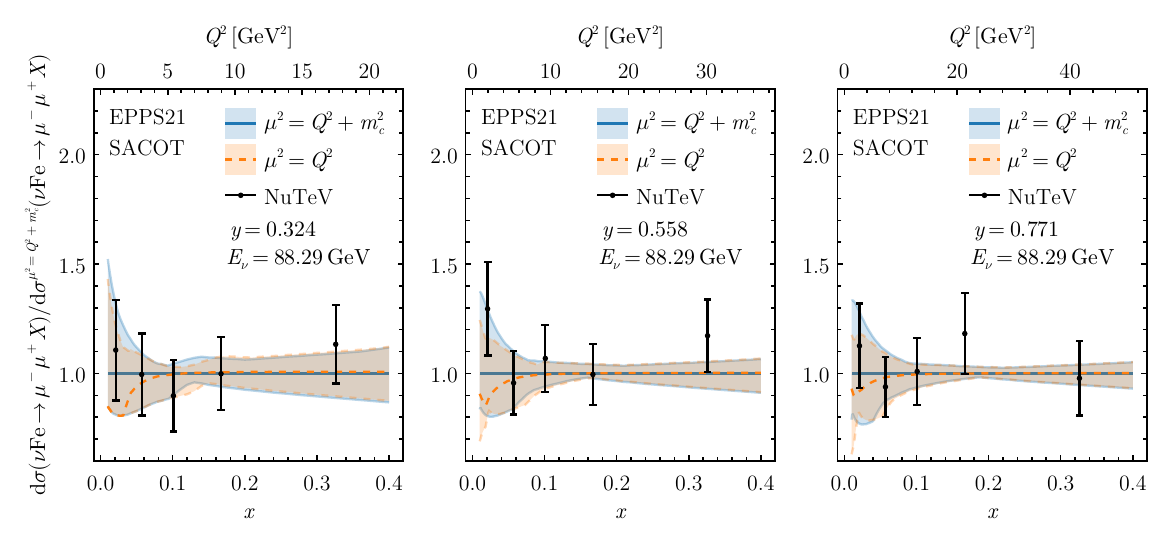}
    \caption{Comparison of the scale choices $\mu^2=Q^2$ and $\mu^2=Q^2+m_c^2$ for dimuon production in neutrino scattering in NuTeV kinematics, computed with \texttt{EPPS21} at NLO using the SR (top row) and SACOT (bottom row) schemes. The uncertainty bands depict the envelope of all 17 scale choice combinations for each central curve. The theoretical calculations, as well as the experimental NuTeV data, are normalized to the central value with $\mu^2=Q^2+m_c^2$.}
    \label{fig:scale_choice_comparison}
\end{figure}

In the massless SIDIS calculation, it is natural to use the scale choice $\mu^2=Q^2$ for all scales (renormalization, factorization, and fragmentation) as the scale logarithms are of the form $\log(Q^2/\mu^2)$. In the case of massive hard coefficients, the scale logarithms look like $\log\left((Q^2+m_c^2)/\mu^2\right)$, making the choice $\mu^2=Q^2+m_c^2$ more natural. Another benefit of this particular choice is that the scale will always be above the charm-mass threshold (except when multiplied by a factor $1/4$ for the estimation of scale uncertainties). The differences between these two choices are illustrated in figure~\ref{fig:scale_choice_comparison}. The overall behaviour is the same in both the SR and the SACOT calculations: at small values of $Q^2$, the central cross-section value with $\mu^2=Q^2$ falls \SIrange{10}{20}{\percent} more sharply in comparison to the central value with $\mu^2=Q^2+m_c^2$, as the strange-quark distribution is an increasing function of factorization scale at small $x$. However, this difference is captured well by the scale uncertainty, as the scale envelope remains the same in both cases. As is to be expected with all charm-mass effects, the differences start to diminish as $Q^2$ grows. 

Below the charm-mass threshold $Q^2<m_c^2$, all three scales are fixed to the charm mass $m_c^2$, which is determined from the PDF unless otherwise stated. This is because typically PDFs are not evolved below the charm-mass threshold. The FFs, on the other hand, become scale-independent below this threshold. Moving forward, we will use the scale choice
\begin{equation}
    \mu^2=\max\left\{k^2\left(Q^2+m_c^2\right), m_c^2\right\}; \quad k\in \{\tfrac{1}{2}, 1, 2\}
\end{equation}
for all calculations, unless explicitly indicated otherwise. The central values use $k=1$, while the scale uncertainty estimation uses also $k=\frac{1}{2}, 2$. The prescription for obtaining the estimate on the scale uncertainty is to compute the cross section with all three choices of $k$ for all three scales with the restriction
\begin{equation}
    \label{eq:scale_variation_bounds}
    \frac{1}{2}\renorm\leq \fact, \frag\leq 2\renorm.
\end{equation}
This restriction removes cases that would give excessively large scale logarithms, resulting in a total of 17 different scale choices. The 17-point scale uncertainty is then simply the envelope of all these scale choices. It should be noted that the absolute values of the scale uncertainty do not have a clear interpretation, but the scale uncertainties give an idea of how sensitive each calculation is to the scale choice and thus to the missing higher-order corrections.

\begin{figure}
    \centering
    \includegraphics[width=\linewidth]{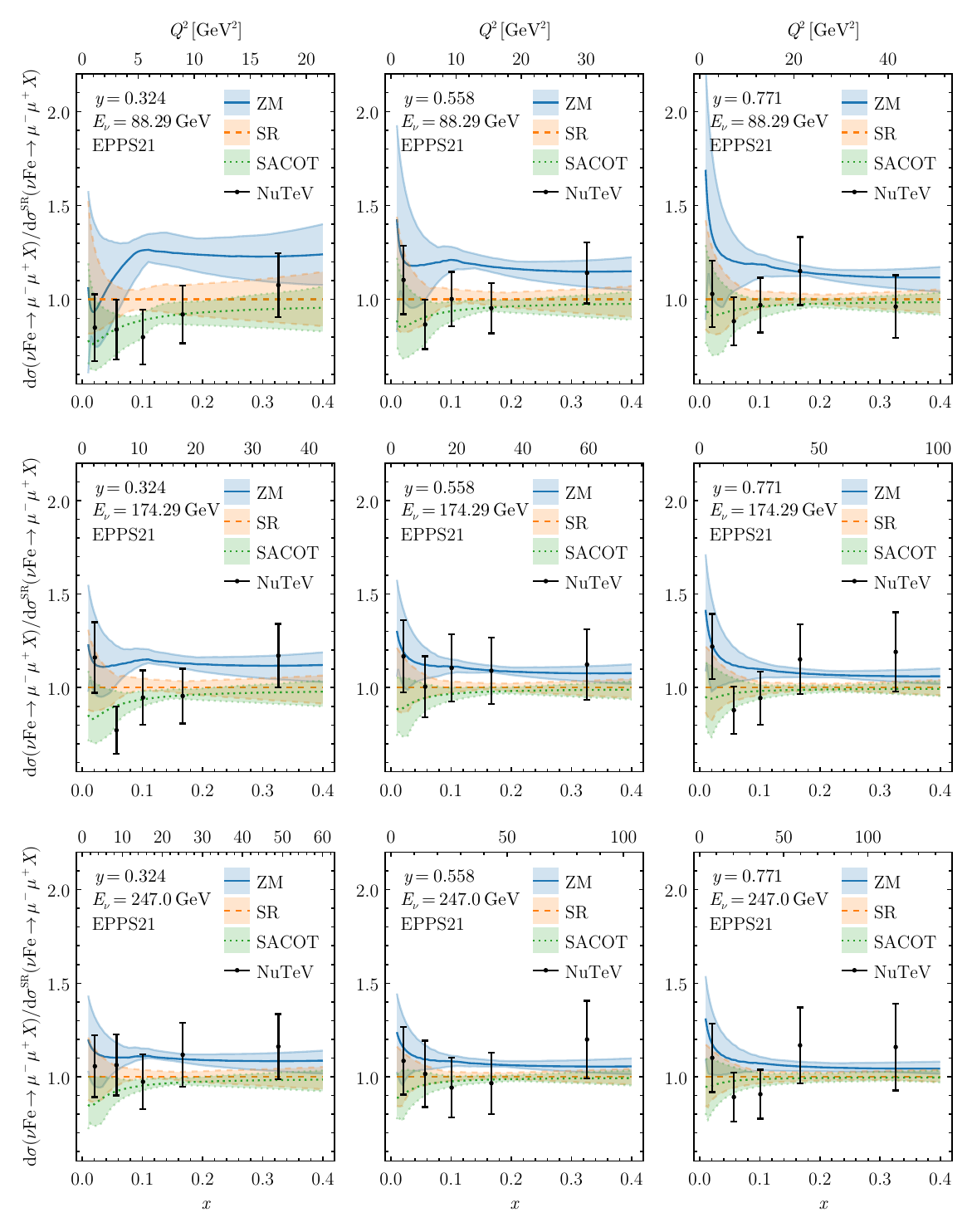}
    \caption{Comparison of the ZM, SR, and SACOT schemes for dimuon production in neutrino scattering in the NuTeV kinematics, computed with \texttt{EPPS21} at NLO using the scale choice $\mu^2=Q^2+m_c^2$. The uncertainty bands depict the envelope of all 17 scale choice combinations for each central curve. The theoretical calculations, as well as the experimental NuTeV data, are normalized to the central SR value.}
    \label{fig:scheme_comparison_neutrino}
\end{figure}

\begin{figure}
    \centering
    \includegraphics[width=\linewidth]{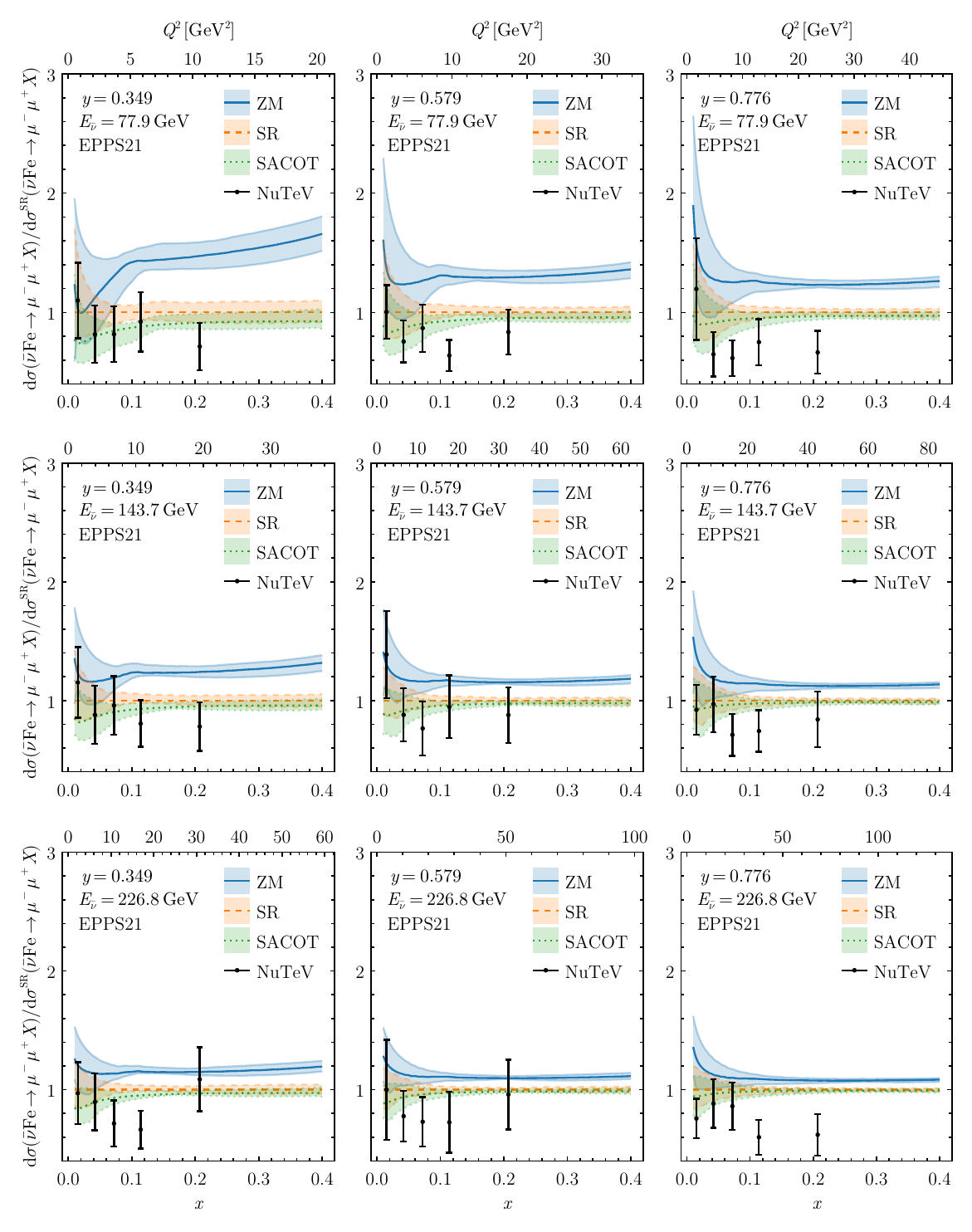}
    \caption{Same as figure~\ref{fig:scheme_comparison_neutrino}, but for antineutrino scattering.}
    \label{fig:scheme_comparison_antineutrino}
\end{figure}

In figures \ref{fig:scheme_comparison_neutrino} and \ref{fig:scheme_comparison_antineutrino}, we show the differences between the zero-mass (ZM), slow-rescaling (SR), and SACOT schemes. All three calculations use the same scales ($\mu^2=Q^2+m_c^2$). Both the ZM and SR schemes use massless coefficient functions, with the difference being that only the SR scheme uses the slow-rescaling variable $\chi$ in the convolutions whereas the ZM scheme uses the plain momentum fraction $x$. As is to be expected, the differences between these three are most significant at small $Q^2$ and start to diminish as $Q^2$ grows. The ZM cross section is consistently larger than the SR and SACOT ones. The introduction of the slow-rescaling variable when moving from the ZM calculation to the SR one has two effects. Firstly, it shifts the argument of the PDF from e.g. $q(x, \fact^2)$ to $q(\chi, \fact^2)$. Since the strange-quark distribution is decreasing as a function of the momentum fraction, the shift from $x$ to $\chi>x$ will decrease the PDF and consequently the cross-section value. Secondly, it restricts the available phase space by changing the lower integration limit of the convolution from $x$ to $\chi$. Thirdly, it scales the normalization factor $\kappa_2$ of $F_2$ (see eq.~\eqref{eq:sidis_normalizations}) from $2x$ to $2\chi$. Of these effects, the shift in the PDF momentum fraction argument is expected to be the most significant. The inclusion of matrix-element level mass corrections then further decreases the cross section value. Depending on the kinematic region, the effect of going from the SR scheme to the SACOT scheme can have an effect from a few percent up to $\SI{20}{\percent}$, again being more important at small values of $Q^2$.

\begin{figure}
    \centering
    \includegraphics[width=\linewidth]{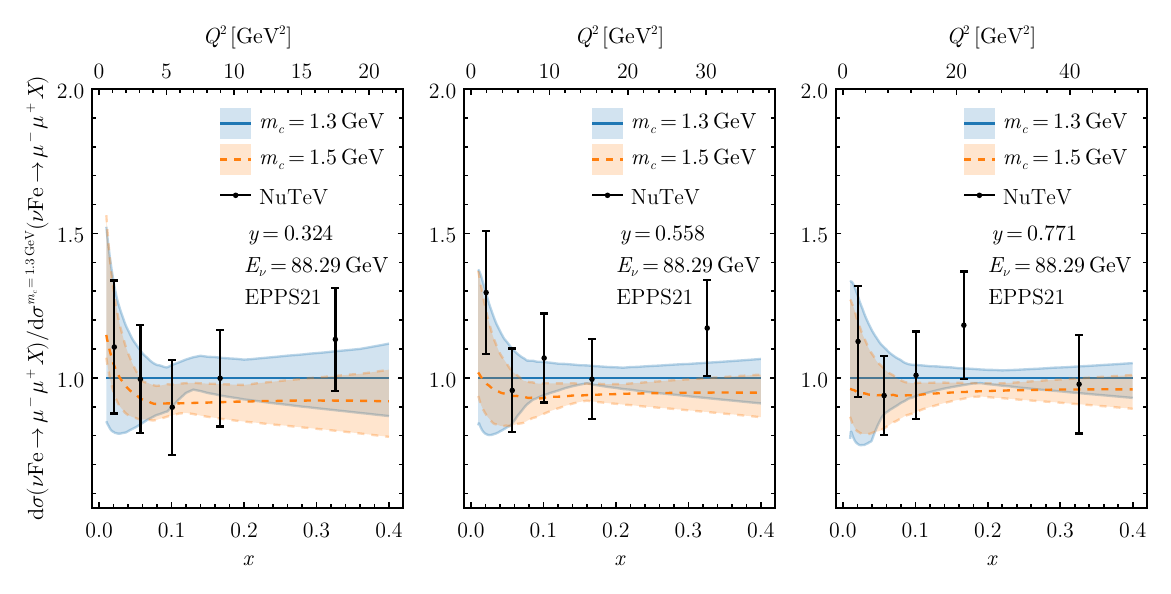}
    \includegraphics[width=\linewidth]{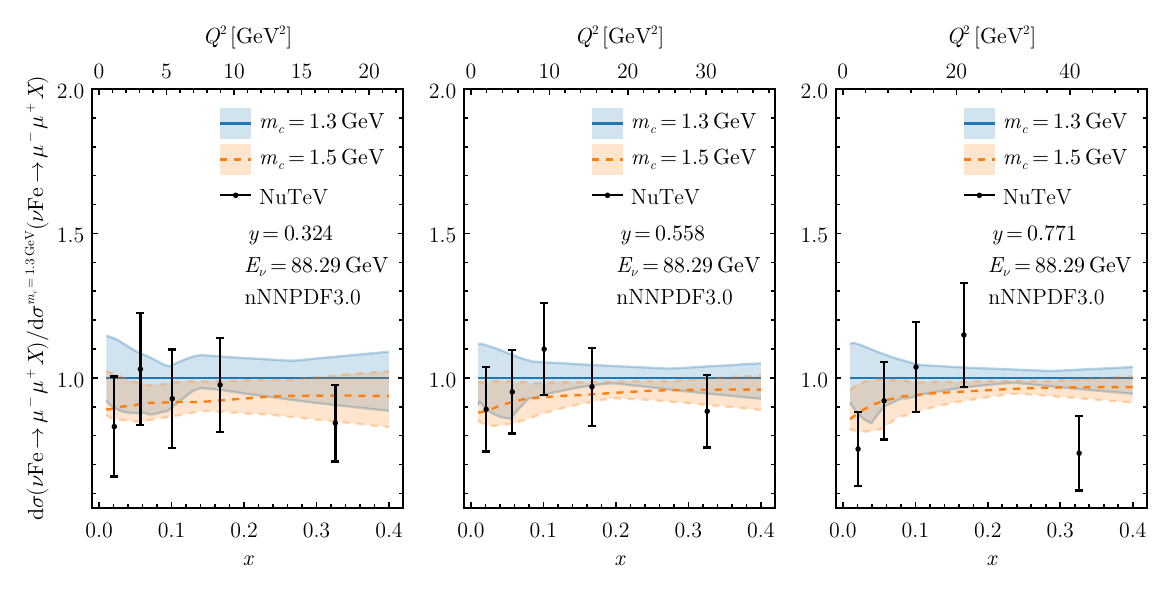}
    \caption{Comparison of the charm-quark masses $m_c=\SI{1.3}{GeV}$ and $m_c=\SI{1.5}{GeV}$ for dimuon production in neutrino scattering in the NuTeV kinematics, computed with \texttt{EPPS21} (top row) and \texttt{nNNPDF3.0} (bottom row) at NLO using the scale choice $\mu^2=Q^2+m_c^2$. The uncertainty bands depict the envelope of all 17 scale choice combinations for each central curve. The theoretical calculations, as well as the experimental NuTeV data, are normalized to the central value with $m_c=\SI{1.3}{GeV}$.}
    \label{fig:mass_comparison}
\end{figure}

The choice of charm mass is somewhat ambiguous. By default, we use the charm mass as in the given PDF fit. For the \texttt{EPPS21} and \texttt{nCTEQ15HQ} sets this is $m_c=\SI{1.3}{GeV}$ and for \texttt{nNNPDF3.0} it is $m_c=\SI{1.5}{GeV}$.\footnote{The charm-mass value in the \texttt{nCTEQ15HQ} set is $m_c=\SI{1.3}{GeV}$, while the minimum-$Q$ of the LHAPDF \cite{Buckley:2014ana} grid is $Q_{\text{min}}=\SI{1.3001}{GeV}$. We thus use a value of $m_c=\SI{1.30011}{GeV}$ in the calculation for \texttt{nCTEQ15HQ} to prevent the calculation from going below $Q_{\text{min}}$. As the cross section is a continuous function of $m_c$, the difference between $m_c=\SI{1.3}{GeV}$ and $m_c=\SI{1.30011}{GeV}$ is completely negligble.} However, the FF sets use the charm mass $m_c=\SI{1.5}{GeV}$. Thus, when using \texttt{EPPS21} and \texttt{nCTEQ15HQ}, there is a mismatch between the charm-mass values. Figure~\ref{fig:mass_comparison} shows the impact of choosing a different charm mass for the \texttt{EPPS21} and \texttt{nNNPDF3.0} sets. The set \texttt{nCTEQ15HQ} behaves analogously to \texttt{EPPS21}, but \texttt{nNNPDF3.0} is shown separately as this set uses the value $m_c=\SI{1.5}{GeV}$, which matches the FF value. It should be noted that changing the charm-mass value means changing it everywhere in the calculation: in the scales $\mu^2=Q^2+m_c^2$, in the slow-rescaling variable $\chi$, and in the hard coefficients $H_i$. We observe a decrease of up to $\SI{10}{\percent}$ in the cross section going from $m_c=\SI{1.3}{GeV}$ to $m_c=\SI{1.5}{GeV}$. 

\begin{figure}
    \centering
    \includegraphics[width=\linewidth]{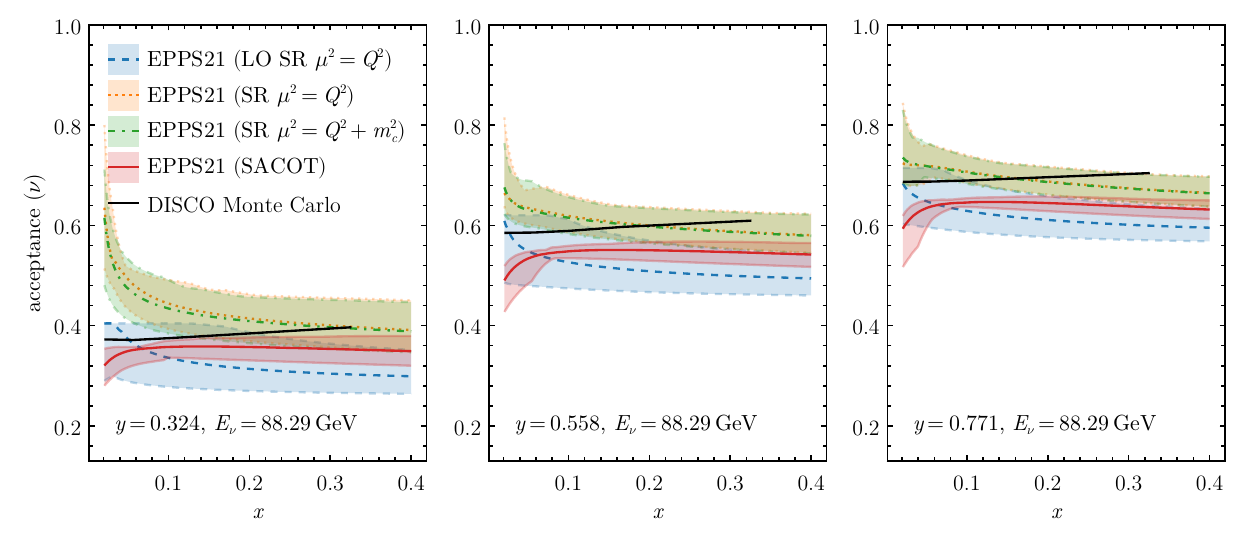}
    \caption{Acceptance correction, as defined in eq.~\eqref{eq:acceptance_def}, in neutrino scattering and in the NuTeV kinematics evaluated using \texttt{EPPS21} in the SR and SACOT schemes at LO and NLO. The SR-scheme calculation uses both scale choices $\mu^2=Q^2$ and $\mu^2=Q^2+m_c^2$, while the SACOT-scheme calculation is only for $\mu^2=Q^2+m_c^2$. The uncertainty bands correspond to the 17-point scale envelopes. Our calculation is compared against the DISCO Monte-Carlo calculation from ref.~\cite{Mason:2006qa}.}
    \label{fig:acceptance_scheme_comparison}
\end{figure}

\begin{figure}
    \centering
    \begin{subfigure}{0.38\textwidth}
        \includegraphics[width=\linewidth]{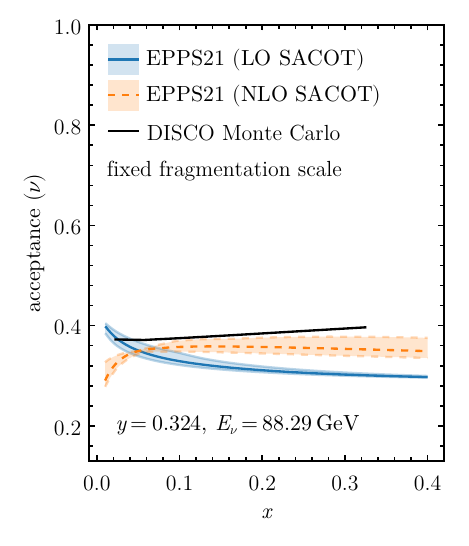}
        \caption{With charm-initiated channels} 
        \label{fig:acceptance_7p_vfns}
    \end{subfigure}%
    \quad
    \begin{subfigure}{0.38\textwidth}
        \includegraphics[width=\linewidth]{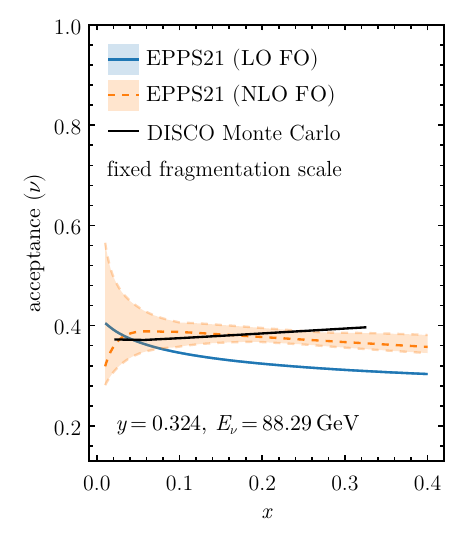}
        \caption{Without charm-initiated channels} 
        \label{fig:acceptance_7p_ffns}
    \end{subfigure}
    \caption{Acceptance correction, as defined in eq.~\eqref{eq:acceptance_def}, in neutrino scattering and in the NuTeV kinematics evaluated using \texttt{EPPS21} in the SACOT scheme at LO and NLO. The uncertainty bands correspond to the 7-point scale envelopes (i.e. the fragmentation scale is fixed to $\frag^2=Q^2+m_c^2$). Our calculation is compared against the DISCO Monte-Carlo calculation from ref.~\cite{Mason:2006qa}. The left-hand panel includes the charm-initiated channels in the inclusive DIS calculation, whereas the right-hand panel does not include these channels.}
    \label{fig:acceptance_7p}
\end{figure}

\begin{figure}
    \centering
    \includegraphics[width=\linewidth]{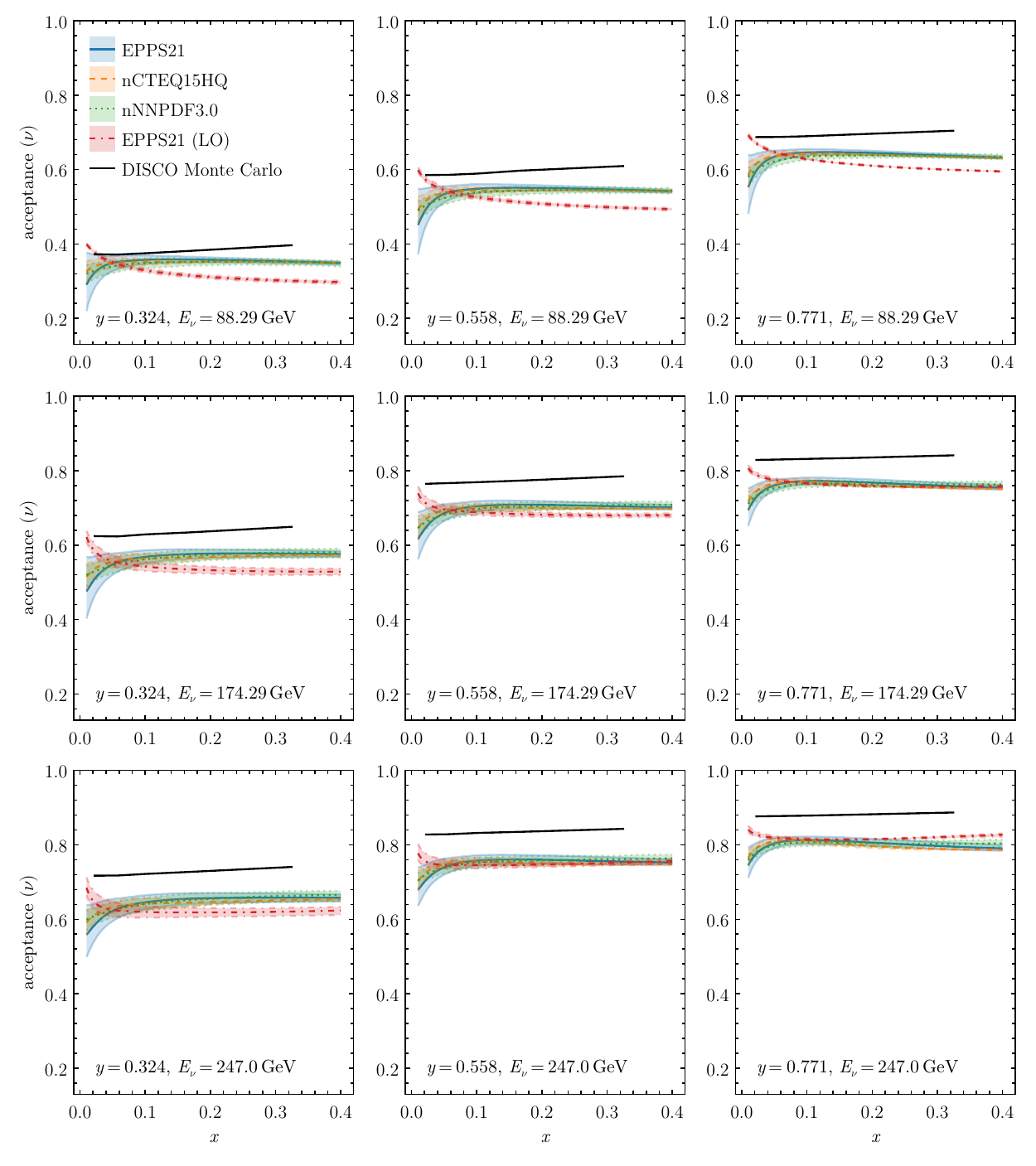}
    \caption{Acceptance correction, as defined in eq.~\eqref{eq:acceptance_def}, in neutrino scattering and in the NuTeV kinematics evaluated using the \texttt{EPPS21}, \texttt{nCTEQ15HQ}, and \texttt{nNNPDF3.0} sets in the SACOT scheme at NLO using the scale choice $\mu^2=Q^2+m_c^2$. The uncertainty bands correspond to the $\SI{90}{\percent}$ confidence-level PDF uncertainties. Our calculation is compared against the DISCO Monte-Carlo calculation from ref.~\cite{Mason:2006qa}.}
    \label{fig:acceptance_neutrino}
\end{figure}

\begin{figure}
    \centering
    \includegraphics[width=\linewidth]{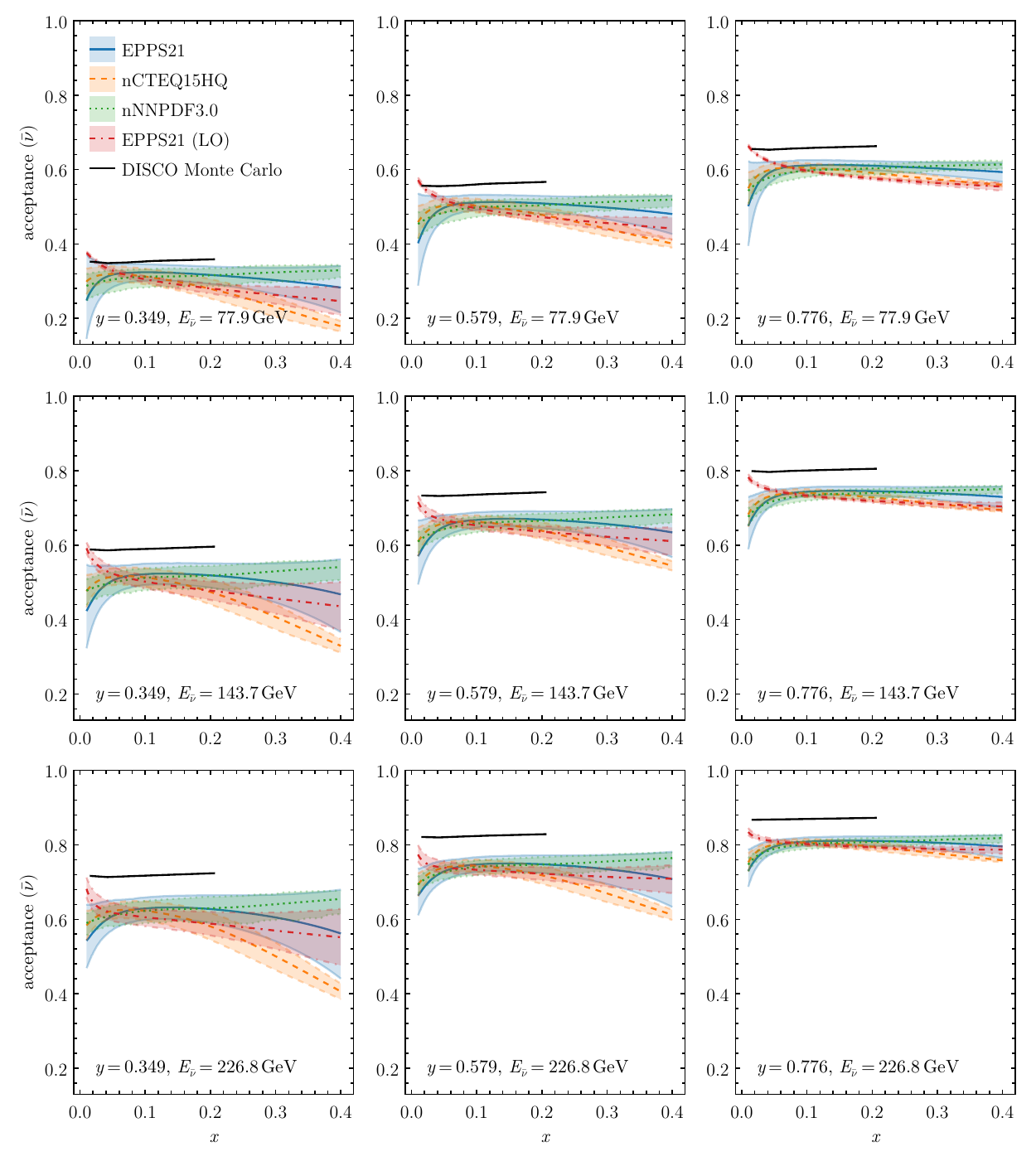}
    \caption{Same as in figure~\ref{fig:acceptance_neutrino}, but for antineutrino scattering.}
    \label{fig:acceptance_antineutrino}
\end{figure}

The acceptance correction, which is usually taken as input, is a derived quantity in our framework. We define it as the ratio of the semi-inclusive dimuon and inclusive charm production cross sections (see eq.~\eqref{eq:charm_prod_factorization}):
\begin{equation}
\label{eq:acceptance_def}
    \mathcal{A}=\frac{1}{\mathcal{B}_\mu}\frac{\dd\sigma(\nu_\mu N\to \mu\mu X)}{\dd\sigma(\nu_\mu N\to c\mu X)},
\end{equation}
with $\mathcal{B}_\mu=0.0919 \pm 10\, \%$ \cite{Bolton:1997pq}. Figures \ref{fig:acceptance_scheme_comparison}--\ref{fig:acceptance_antineutrino} show a comparison of the acceptance correction computed in our framework to that computed with the DISCO Monte Carlo framework \cite{Mason:2006qa}. The DISCO-based acceptance is defined as the ratio of generated dimuon events satisfying the decay-muon energy cut to all generated dimuon events. The events are generated according to a fixed-order NLO SIDIS cross section, where the charm fragmentation is modeled with the scale-independent Collins-Spiller form \cite{Collins:1984ms}. The decay channel of the charmed hadron is chosen randomly according to PDG branching fractions, and the kinematics are sampled based on matrix elements calculated in ref.~\cite{Barger:1976ac}. These matrix elements are obtained with simplifying assumptions, such as by using free-quark decay matrices with physical hadron masses.

Figure~\ref{fig:acceptance_scheme_comparison} showcases the scale dependence and the impact of the different schemes to the computed acceptance correction. The choice of scale and scheme is always the same for the semi-inclusive and inclusive cross sections required in eq.~\eqref{eq:acceptance_def}. The SR scheme calculation with $\mu^2=Q^2$ corresponds to our previous work in ref.~\cite{Helenius:2024fow} and acts as a comparison baseline.\footnote{There is a slight difference between the SR scheme calculation in this work and in ref.~\cite{Helenius:2024fow}: in ref.~\cite{Helenius:2024fow}, the inclusive $F_2$ was taken to be proportional to $\chi$ for the induced LO channels $\overline c\to \overline d/\overline s/\overline b$. As explained in section~\ref{sec:inclusive_dis_sacot}, we now take the $F_2$ for the induced channels to be proportional to only $x$. The difference for the acceptance correction, however, is negligble.} The difference between the scale choices $\mu^2=Q^2$ and $\mu^2=Q^2+m_c^2$ is small, but the scale uncertainty itself is considerable. There is also a significant difference going from the SR scheme to the SACOT scheme (with the scale choice $\mu^2=Q^2+m_c^2$). The acceptance in the SACOT scheme is routinely $\SIrange{5}{15}{\percent}$ below the SR scheme. At the smallest values of $Q^2$, this difference can be as large as $\SI{100}{\percent}$, although the artifacts from the choice of how to handle the regime below the charm-mass threshold likely dominate the difference. However, this only underlines the volatility of the acceptance correction and its significant scheme-dependence.

The scale envelope in figure~\ref{fig:acceptance_scheme_comparison} is obtained with the 17-point method, i.e. all three scales are varied independently within the bounds established in eq.~\eqref{eq:scale_variation_bounds}. For the inclusive case this means that some of the scale variations will produce identical results, as there is no fragmentation scale in the inclusive calculation. With the 17-point method, we capture the effects of varying the fragmentation scale. It is precisely this variation of the fragmentation scale that results in the large scale dependence on the LO calculation. Using instead the 7-point method, where the fragmentation scale is fixed, the scale dependence in the LO case goes down significantly, as shown in fig.~\ref{fig:acceptance_7p_vfns}. Some scale dependence is still left due to the induced charm-initiated channels in eq.~\eqref{eq:inclusive_dis_sacot}. This is precisely the statement that the factorization in eq.~\eqref{eq:charm_prod_factorization} breaks down even for the LO VFNS. The scale dependence vanishes completely only if the LO calculation is done in a FFNS, which is demonstrated in fig.~\ref{fig:acceptance_7p_ffns}.

Figures \ref{fig:acceptance_neutrino} and \ref{fig:acceptance_antineutrino} then quantify the PDF-dependence of the acceptance correction. Particularly in the case of antineutrino scattering (figure \ref{fig:acceptance_antineutrino}), the differences between the three PDF sets, as well as the propagated uncertainties, are sizeable. The systematic differences between our calculation and the DISCO Monte Carlo calculation is also noteworthy. While the bulk of the differences between these calculations can be compensated with the $\SI{10}{\percent}$ normalization uncertainty associated with the semi-leptonic branching ratio $\mathcal{B}_\mu$, some kinematics-dependent differences remain. This is most noticable when comparing the top-left panel (with the smallest inelasticity and beam energy) to the other panels. In the top-left panels, the two calculations agree well at the smallest values of $x$ and start to slowly diverge as $x$ grows.

\begin{figure}
    \centering
    \includegraphics[width=\linewidth]{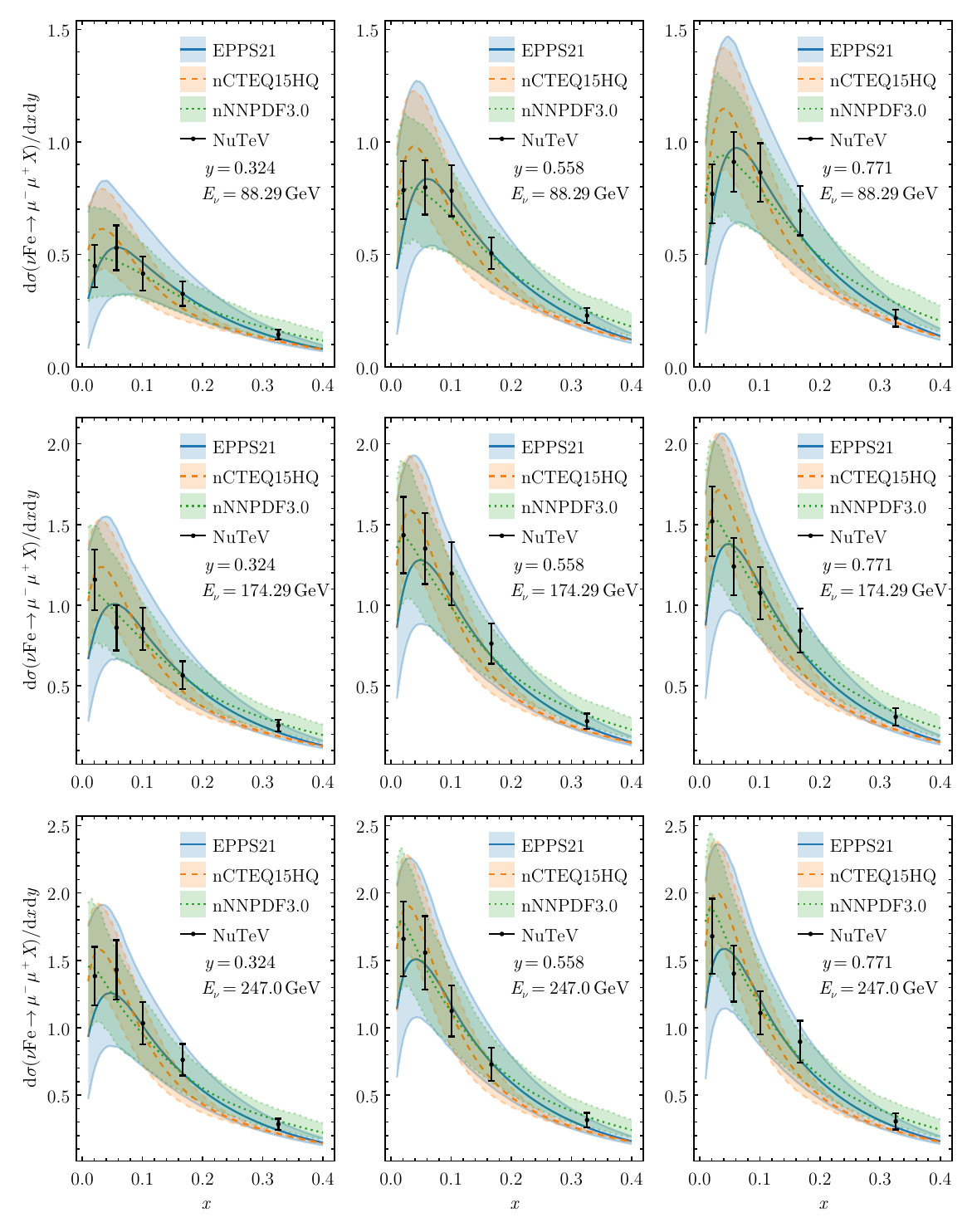}
    \caption{Neutrino dimuon cross sections in the NuTeV kinematics evaluated using the \texttt{EPPS21}, \texttt{nCTEQ15HQ}, and \texttt{nNNPDF3.0} PDF sets in the SACOT scheme at NLO using the scale choice $\mu^2=Q^2+m_c^2$. The uncertainty bands depict the $\SI{90}{\percent}$ confidence-level PDF uncertainties. The theoretical calculations are compared against NuTeV data \cite{NuTeV:2007uwm}. The cross-section values should be multiplied by $G_F^2 M E_\nu/100\pi$.}
    \label{fig:nutev_neutrino}
\end{figure}

\begin{figure}
    \centering
    \includegraphics[width=\linewidth]{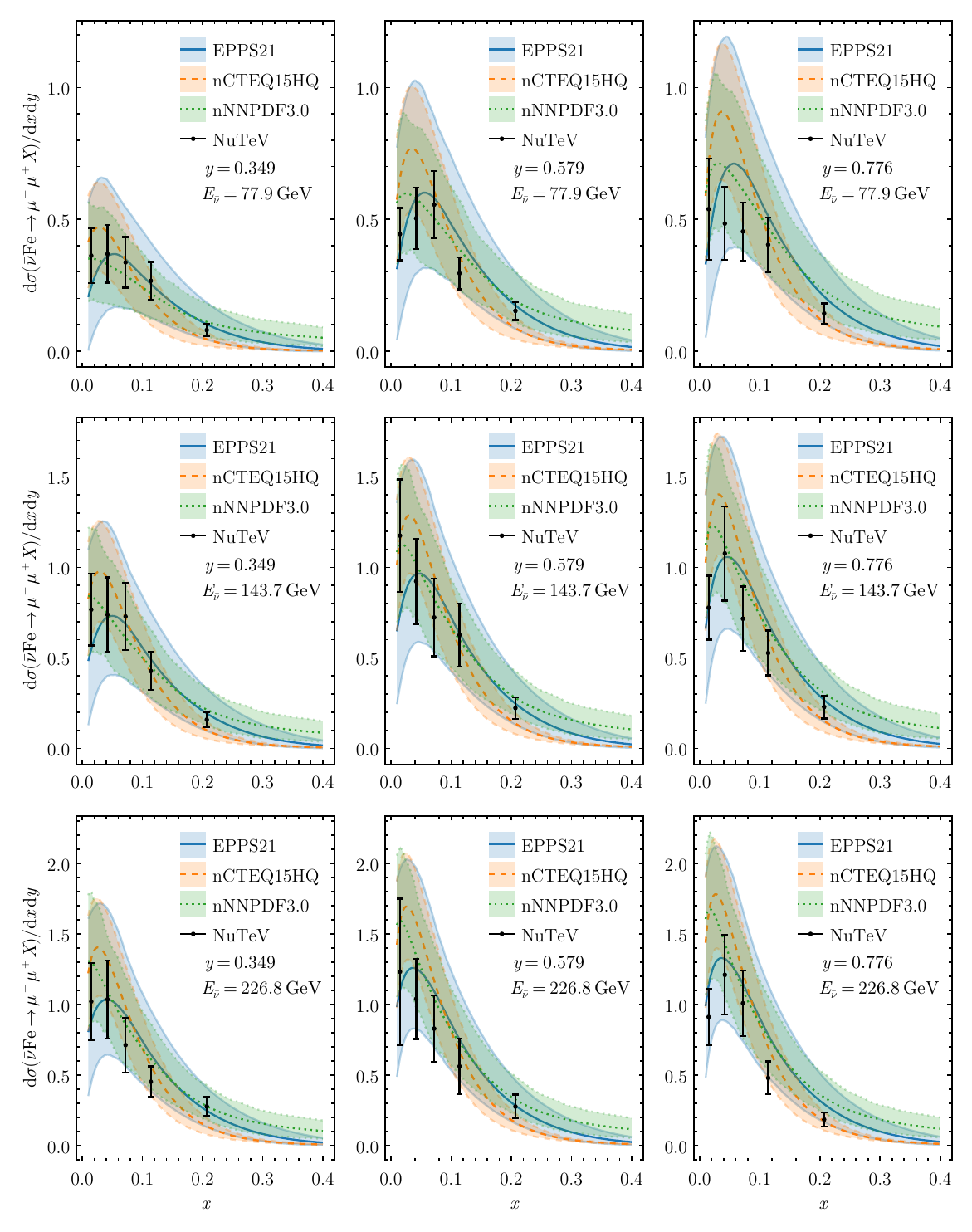}
    \caption{Same as figure~\ref{fig:nutev_neutrino}, but for antineutrino scattering.}
    \label{fig:nutev_antineutrino}
\end{figure}

Finally, we present the full dimuon production cross sections in the SACOT-$\chi$ scheme in figures \ref{fig:nutev_neutrino} and \ref{fig:nutev_antineutrino}. As in our previous work, these results demonstrate that we can describe pre-existing NuTeV data well in our framework. The systematics between the three PDF sets are similar in the SACOT scheme presented here as in the SR scheme presented in ref.~\cite{Helenius:2024fow}. Most importantly, the systematic differences between the sets still reflect the differences in the underlying strange-quark distributions. The agreement with the data remains similar, even if the agreement for any given individual data point might differ when going from the SR calculation to the SACOT one. The PDF uncertainties remain significant, indicating potential for better constraints from these data. As the full mass corrections have the tendency to reduce the cross section values, performing a fit without these corrections could lead to an underestimation of the strange and particularly the antistrange distributions.

\section{Conclusion and outlook}
\label{sec:conclusion}

In this paper, we have presented a self-contained NLO GM-VFNS calculation of dimuon production in neutrino-nucleus collisions in the SACOT-$\chi$ scheme. Our previous framework, which only included kinematic charm-mass effects, improved existing calculations found in literature by foregoing the approximative factorization to inclusive charm production and instead using DGLAP-evolved fragmentation functions and a muonic decay function, both fitted to experimental data. 

In our main result, we considered dimuon production in three different mass schemes (ZM, SR, and SACOT-$\chi$). We found the results to align with our expectations: the kinematic mass corrections included in the SR scheme are the most significant and the dynamic mass corrections in the SACOT-$\chi$ scheme are subleading. Furthermore, we found the SR and SACOT-$\chi$ calculations to convergence in the limit of large $Q^2$, as is to be expected. Overall, the dynamic mass corrections in the SACOT-$\chi$ calculation introduce an up-to $\SI{20}{\percent}$ correction to the SR calculation, depending on the kinematic region. Thus, a heavy-quark scheme is required for the low-$Q^2$ kinematic region, which then allows for a consistent inclusion of such dimuon data to a (nuclear) PDF analysis. The considered data clearly offer potential for further constraining the strange distribution in a nuclear PDF fit. However, as most nuclear PDF fits use a proton PDF as a baseline against which nuclear corrections are fitted, care must be taken if the same data have already been included in the baseline fit.

We also considered the impact of the choice of charm mass, as it is typically taken as a fixed parameter in PDF and FF fits, with different fits using different values. The typical mass values are $m_c=\SI{1.3}{GeV}$ and $\SI{1.5}{GeV}$, which are also the ones we compared. The impact of this choice is between a few percent to up to $\SI{10}{\percent}$ in NuTeV kinematics, depending on the exact kinematic region.

We reiterated the dependence of the effective acceptance correction on the PDFs and perturbative order, and now showed that it also depends on the scales and the scheme.\footnote{Numerical tables of the acceptance values are available as ancilliary files in the arXiv preprint repository.} The fact that the acceptance correction --- a necessary input in pre-existing calculations --- depends on more than just kinematics shows the breakdown of the approximate factorization assumption. This makes our calculation the first fully consistent GM-VFNS implementation of dimuon production.

In the future, we plan on implementing the full NNLO corrections in the SR scheme. The massless photon-current NNLO coefficients have recently been made available \cite{Bonino:2024qbh,Bonino:2024wgg,Goyal:2023zdi,Goyal:2024emo,Goyal:2024tmo}, and the charged-current coefficients are expected to be made publicly available in the near future \cite{Bonino:2025tnf}.\footnote{The NNLO charged-current coefficients have become available during the review process of this publication; see ref.~\cite{Bonino:2025qta} and also ref.~\cite{Haug:2025ava}.} As was discussed before, dynamical mass corrections are subleading. Thus, we expect that including only the kinematical mass effects at NNLO will provide an accurate approximation of the full GM-VFNS NNLO calculation. Reducing the scheme uncertainty associated with the charm mass, however, will require the full GM-VFNS calculation. In addition, work is ongoing to study the effects of including NuTeV and CCFR data with the SIDIS-based approach presented in this work to a nuclear PDF fit using reweighting methods. It is also worth pointing out that we cannot currently quantify the uncertainties arising from the fragmentation functions, as the available FFs do not include any error sets. Should charmed-hadron FF fits with error sets become available, we could also study this relevant source of uncertainty.

\section*{Acknowledgements}

We acknowledge the financial support from the Magnus Ehrnrooth foundation (S.Y.), the Research Council of Finland Projects No. 331545 and 361179 (I.H.), and the Center of Excellence in Quark Matter of the Research Council of Finland, project 364194. The reported work is associated with the European Research Council project ERC-2018-ADG-835105 YoctoLHC. We acknowledge grants of computer capacity from the Finnish IT Center for Science (CSC), under the project jyy2580. 

\bibliographystyle{JHEP}
\bibliography{refs.bib}

\end{document}